# AN ANALTICAL OVERVIEW OF VIRTUAL MACHINE LOAD BALANCING SCHEDULING ALGORITHMS: A COMPARATIVE STUDY OF PSO, MOO AND ACTIVE MONITORING ALGORITHM

**Priyank Vaidya[1], Murli Patel[1], Abhinav Sharma[1], Dr. Nishant Doshi[1]**

Department of Computer Science and Engineering, Pandit Deendayal Energy University, India.

priyankhvaidya@gmail.com, abhinav291sharma@gmail.com, patelmurli15@gmail.com, Nishant.Doshi@sot.pdpu.ac.in

## ABSTRACT

This paper contains the analysis of VM load balancing scheduling algorithms discussing its advantages, disadvantages along with applications. As the Industry shifts towards adapting cloud technologies, it is important to optimally load balance the client requests to the servers. It becomes important to the cloud providers to adapt technology so that its customers don't fall behind of latency issues. The algorithm which is more often used in the load balancers are analytically discussed. The fundamentals regarding virtualisation & VM allocation concepts are also covered.

**Keywords:** Virtual Machine Scheduling, Load Balancing, Cloud Computing, Virtualization, Particle Swarn, Multi-o, Active Monitoring

## 1.0 INTRODUCTION

Cloud Computing Technologies are new wave in the Tech Market, rising from all industries pushing services to cloud, it is very important for cloud providers to provide secure, reliable, and low latency infrastructure. Cloud Providers providing services to multiple clients across different Industry sectors, often find difficulty in giving cutting edge solutions. Cloud Providers utilise their technology in such manner irrespective of Load on the Servers, it is capable to provide cloud services to clients. For Proper utilisation of resources, it is required for them to efficiently and optimally schedule the load to Virtual Machines. The VMs on top of Virtualisation are the essential part of any Cloud Infrastructure. The algorithms discussed in are ways in which load is balanced in the distributed systems achieving lower latency and quick feedbacks from the underlying deployed infrastructure. The Algorithm are discussed based upon each one's analysis and at the end suggestions along with type of applications which is used. There are two type pf algorithm which are used.

- Static Algorithms: These algorithms often require real time inputs to balance the load. This type of Algorithms is complex and heavy when compared to dynamic algorithms.
- Dynamic Algorithms: These algorithms take decisions based upon the load and does not require real time updates to optimally balance the load. These are largely used in Distributed Systems.

This paper discusses the Dynamic Load Balancing Algorithms, which are used for specific applications in Large Distributed Systems.

## 2.0 VIRTUALISATION TECHNOLOGY

Consider that you want to deploy the different applications on the single hardware. You might think that this can be done by abstracting the application in a single file and run that when in use. This can be quite tedious to do for a normal user who doesn't wants to separate the applications directly on the hardware level. There the concept of Virtualisation comes, which allows to create the virtual environment on the underlying hardware/network and storage resources, i.e., at the same abstraction level.

Virtualisation concepts helps in creating the virtual machine with its operating system and underlying separated resources. The Virtualization has benefits of configuring different application requirements onto the same deployment environment. As we are talking at the level of servers, therefore, host hardware here denotes to the Physical Server.

The separation of the real hardware resources to get the virtualised resources can either happen statically or dynamically. Through Virtualisation it helps the host OS to map the specific/required services to VMs. Load Balancing is the subprocess of virtualisation techniques. Ensuring that efficient optimisation of tasks/schedules on behalf of VM Resource and its optimisation[1].

Due to these advantages, Virtualisation enables Cloud Computing to provide the best services to the users, whether it be performance, Latency, Cost, Effectiveness, and Environmental Sustainability.

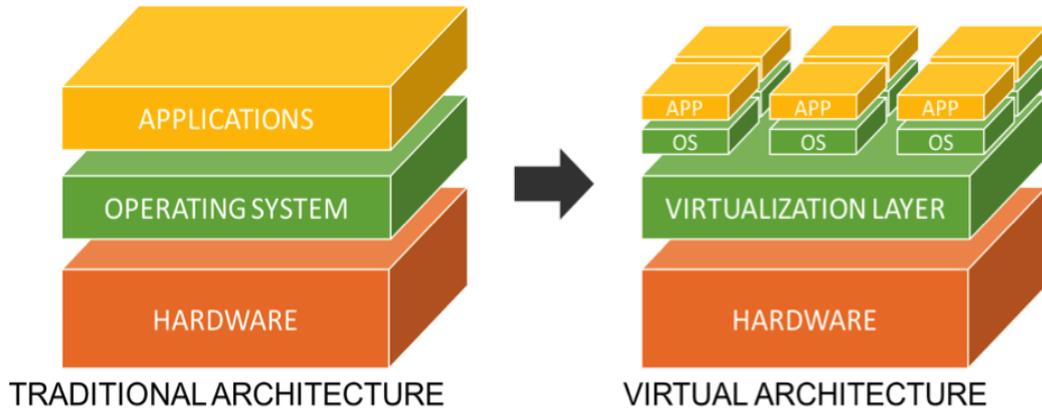

**Figure 1:** Virtualisation Architecture[1].

There might be some concerns regarding the need for Virtualisation, which includes higher optimisation requirements and efficient operations. Virtualisation allows the deploy the applications/Processes in VMs which is user configurable, thereby getting all the flexibility to virtualise the host hardware. Virtualisation performed by Hypervisors (can be Type1/Type2) can be let the VMs to run independently having the gist of real hardware environment apart from the virtualised environment[2, 3].

## 2.0.1 VM ALLOCATION

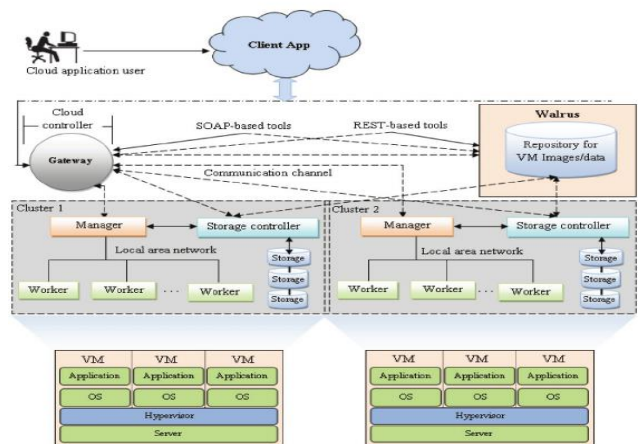

**Figure 2:** *VM Allocation Architecture*[3].

Understanding VM allocation of private cloud networks will let us to understand more about implications of VM Allocation. Private Cloud Architectures like Eucalyptus segments the VM instances into multiple controllers. Depending upon the user/admin requests, the controller is invoked returning the required response Contexts like controller are API driven, which are internally connected that enables provisioning of VM instances, storage, scheduling, and User response[3, 1, 2]. For the VM instance setup take place in cloud controllers which automates the VM allocation, storage, Networking, and Infrastructure on the fundamental level. This makes the effort to create a fundamental for dynamic VM allocation. For dynamic scheduling, the infrastructure needs to set up cloud managers which will improve the cluster and resource pool utilisation with respect to the load of incoming requests. Parallel computing is optimised through such operations at the kernel level of server, giving the flexibility to model the VM mapping. Manager works at the LAN layer, connecting directly to the client gateway, allowing users to query the responses faster, and highly dynamically than normal computing methods[3, 5, 14].

## 2.1 VM SCHEDULING ALGORITHMS FOR LOAD BALANCING

Industry faces major problems while working on operational fields, especially to configure Load balancers. One of the types of Load Balancers is VM Load Balancer. VM Scheduling is the product of Load balancing allowing the application to get deployed if one of the VM gets fault tolerance. When Operational team sets up the VM in cloud, the team has to take care of conditions in the applications which can lead to fault tolerance and also make the decisions to address the issue[4].

For the instance, the Cloud Architects also must frame algorithms which can shift to another VMs if one of the VM gets fault tolerated. Such algorithms will make the application to run smoothly even if some exceptions occur. VM Scheduling Algorithms are highly dependent on Task and Resource Scheduling, as they perform at the fundamental level of any deployed applications. Task distribution is performed by breaking the application into various processes, these requires Resources which are either statically or dynamically allotted to the Processes. Before understanding the method of scheduling of Task and Resource scheduling, it doesn't make sense to understand VM Scheduling algorithms[18, 20].

VM Scheduling is always a 2-Level Scheduling algorithm. 1st Level maps the Guest Operating system's Processes and threads with the Virtual Resources allotted by VMM. 2nd Level (VMM) Maps the Virtual resources of the Guest operating System, with the real physical hardware of the server. Secondly, High Level of Abstraction; In Containerisation, the process and threads are directly managed by Operating System through the Scheduler. Scheduler identifies the Synchronisation points and allow the next process to get ready for execution, leading to performance efficiency[5, 1].

However in the VM Scheduling takes place at high level of abstraction, hence the management of processes is not efficiently performed in VM's[1, 14].

### *2.1.1 Advantages of VM Scheduling Algorithm*

- Ease of Resource Reservation: In certain cases of high computational tasks, the resources need to be reserved for that next instruction to be executed. This can be achieved by per-emption, which requires special service from OS, which invokes the application layer by changing its code which is not desirable. Here the benefit of VM comes, in which the decisions are taken in 2nd Level Scheduling and does not require the application layer to change its code for inordinate.

- VM Elasticity: The way to manage the Job tasks with the technique of co-Scheduling enabling better performance and low latency between the co-related work [1].

- Resource Utilisation: VM Scheduling helps to use maximum of the resources of any VM, this optimises Virtual Environment and reduces waste resources which are often unscheduled.

- Quality Of Service(QoS): Virtual machines working in Virtual environment are separated based upon received appropriate level of service. This makes to run applications with different physical requirements to parallelly compute in same Virtual Environment [15, 19].

## 2.2 PSO ALGORITHM

Particle Swarm Optimization: It is a Metaheuristic algorithm based on Search-space mechanism over parameters we define to apply in our application. The overall position and velocity (here Load parameters) are mutated dynamically referring to global maxima, and particle local extremum. This method allows the problem to get optimized based upon over very large spaces of candidate solution[7]. The PSO behaves exploratory behaviour of searching in broader search-space and exploitative behaviour the local optimum. Each particle represents the optimal solution with its position in search space. Each particle will advance towards speed, influencing whole swarm to approach towards the optimal location after some repetitive iterations.

The particles converge based upon two factors.

- Convergence of sequence of solution to a point base upon local optimal solutions i.e., Local extremum
- Converging based upon the swarm's best position over each iteration by learning[7].

### 2.2.1 Advantages of PSO

- Simple Implementation: Each particle adjusts it's travelling based upon its current position and makes a good choice to get the
- Flexibility: Decreases total execution time of the Workflow tasks and improves Neighbourhood communication through topology helps to find best results based upon resources required by user.
- Robustness: Highly efficient global search algorithm and does not require high parameter tuning
- Parallelism: Concurrent processes can run parallelly enabling faster execution times on multi-core processors.

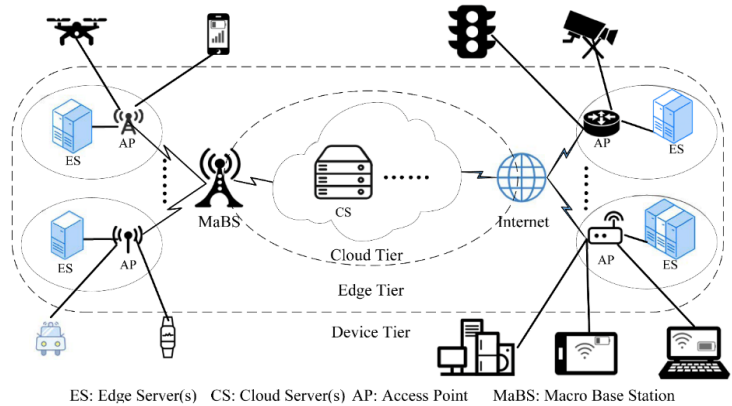

ES: Edge Server(s)   CS: Cloud Server(s)   AP: Access Point     MaBS: Macro Base Station
*Figure 3: PSO based task scheduling [8].*

### 2.2.2 Disadvantages of PSO

- Limited Exploration: Is not optimized for local searchability.

- Stochastic Behaviour: It's convergence ability greatly decreases over time, which makes it non-efficient to use in Higher Dimension space (For larger Load Balancing constraints). Premature Convergence: PSO is prone to premature convergence, which can lead to suboptimal solutions.

- Stochastic Behaviour: The stochastic nature of the algorithm can make it difficult to reproduce results. Limited Exploration: The algorithm may get stuck in local optima and fail to explore the search space adequately.

## 2.2.3 Analysis PSO Algorithm

PSO uses the particle's own information regarding individual and global extremum to guide into next Iteration. Firstly, through Ant Colony Optimization traversing completely will result in local optimal solution and Global optimal solution. Then PSO is used, as speed is hard to express for particles in this context, Cross-Over operation of Genetic Algorithm is used, which let to perform operation on local and global solution producing new location [18]. To Overcome the defect in Algorithm which is falling into local optimal solution for Resource Scheduling. We used ACO and PSO not as competing methods but bring together to find best out of both. The research predicted that this method improved the Resource Utilisation Ratio in the Cloud giving the optimal Scheduling[8].

PSO iteratively communicates with the computer nodes regarding the global optimum state, improving the candidate solution which finds the best optimal node to run the application. PSO applications includes Scheduling in Operating system, as comparatively to Servers, the number of constraints for decision is less. Also, if system uses different Instances, it separates the processes through Type-1 Hypervisor in Kernel enabling user seamless workflow of application.

## 2.3 ACTIVE MONITORING ALGORITHM

### 2.3.1 Introduction

For better optimal VM Scheduling, Active Monitoring helps to schedule the tasks to VM based upon Some Algorithm (Round Robin in this case). Instead checking the availability of VM on VMs level will bring non-optimisable solution. Instead, the Concept in Active Monitoring consists of Data Centre Controller, which keeps track of current task scheduling in VMs based upon each one's response time[9]. Response time being the most important parameter, DCC will take decisions based upon this. If all the VMs are occupied when request comes, new VM is requested to Hypervisor and gets scheduled.

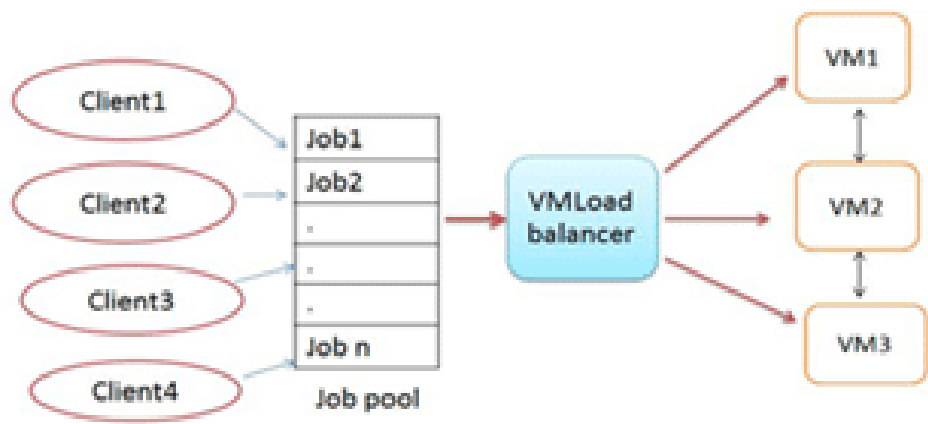

**Figure 4:** *Active Monitoring Scheduler* [9].

The Monitoring is proposed on Load Balancer based on Index table consisting of parameters:
- *VM Primary ID*
- *Current Allocated* load and it's total response time.
- *State of VM:* What state is Server in (Depends on its working condition)
- *Total Offset time:* The total time in Server Life that fails due to high load/request preservation

### 2.3.1 Advantages

- DCC receives the userbase request consisting of the information of request frequency, data size of each request, it's average peak hours (on & off).

- These parameters transcend to the Load Balancer to Index table. Load Balancers go through VM Index table and identifies the VM with less load. If the Load balancer finds more than 1 VM, then load balancer give highest priority to least load[10, 9].

### *2.3.2 Disadvantages*

- Overhead: This algorithm makes significant overheads as it constantly monitors system based upon real-time data. This consumes more resources for computation.
- Complexity: This Algorithm is more complex as it requires constant monitoring to changing conditions. This makes it difficult to maintain.
- Network Latency: Active monitoring techniques could necessitate continuous connection between the monitoring system and the VMs, which could raise network latency and slow the system down[12].

### *2.3.3 Analysis*

The Active monitoring Load balancing algorithm can reduce the response time with maximum VM Resource utilisation. If LB is not properly optimized, it might affect the performance of Cloud and resource underutilisation. Improving LB Algorithms schedules the task in much more efficient way taking into consideration Current load, priority, VM State, and resource availability[10].

The LB Algo can be increases drastically by increasing its computing parameters, which will however bring more efficiency in Decision making for VM/Task Scheduling[22].

## 2.4 MULTI-OBJECTIVE OPTIMISATION ALGORITHM

### 2.4.1 Introduction

There are already some research going in elasticity of cloud. Optimising the VMs considering the dynamicity of cloud is still rare. However, for the specific workload environments, cloud algorithms can be made optimised by minimising the number of VM instances running on server. Considering the cost of resources and load balancing a greater number of VMs will lead to less efficient and highly costly environment. Therefore, utilising the resources highly, by placing the tasks in VMs with faster start up rates brings more efficiency in LB Environments. The optimisation requires relevant fields in any VM with including its load, response time, start-up rate, etc. Bringing all the possibilities in single plate will optimise the VMs load balancing based upon Multi-Objective Optimisation Algorithm[4].

Multi-Objective Optimisation granularizes the task into threads with having similar time to execute (except for certain functions), this brings microservice architecture in the context, bringing all the best fit optimisation methods. Multi-Objective takes care of all the factors considering the resources allocated to the VM and load on it, which makes it to take quick decisions to traffic the incoming request in other VMs or not. Certain Instances like AWS EC2 will take care of Load balancing itself, so that developers do not need to worry to deploy the same application on different instance with the same configuration as previous[4].

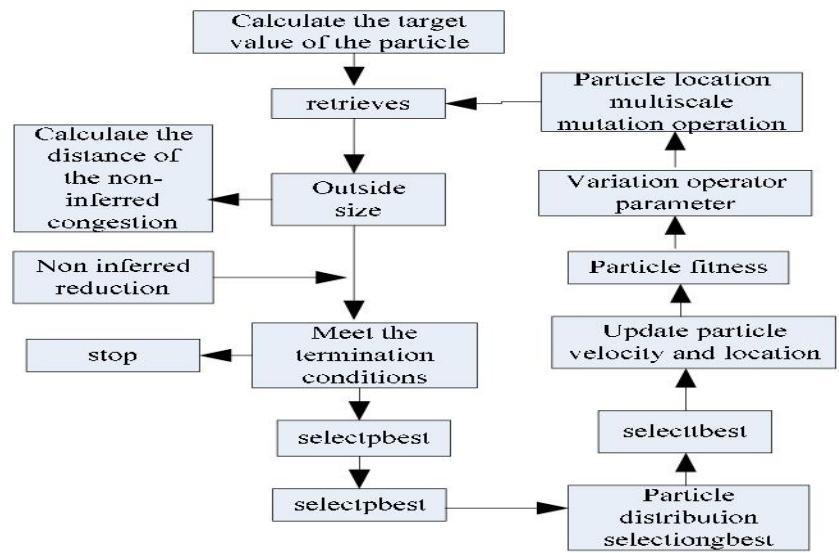

**Figure 5:** *Modelling Analysis and Cost-performance Ratio Optimization of Virtual Machine Scheduling in Cloud Computing* [4].

### 2.4.2 Advantages

- Offers a Pareto Solution: A set of Pareto-optimal solutions may be produced using multi-objective optimization algorithms, enabling decision-makers to select the best one for their requirements and preferences.
- Multi-objective optimization algorithms are built to handle numerous objectives and trade-offs between them, making them suitable for issues where several objectives must be considered.
- Better decision-making: multi-objective optimization can give decision-makers a more thorough and realistic understanding of an issue, enabling them to make more well-informed choices.
- The ability to handle uncertainty can make some multi-objective optimization methods more resilient to changes in the issue or the environment.
- Multi-objective optimization is better able to tackle real-world issues, where there are frequently numerous objectives and trade-offs to be made.

### 2.4.3 Analysis.

- Prior to employing a multi-objective optimization technique, clearly identify the objectives and constraints of the task.
- Apply the proper algorithms: Select an algorithm that is suitable for the aims and problem at hand.
- Certain issues or goals are better suited to algorithms than others.
- Handling constraints correctly is important since some multi-objective optimization techniques may have trouble with them.
- Use parallel computing: Especially for big or high-dimensional problems, use parallel computing to accelerate the optimization process.
- Analyse the answers: Choose the algorithm's answer that best satisfies the decision-tastes maker's and needs after carefully evaluating each one.
- Utilize visualisation: Apply visualisation strategies to better comprehend the trade-offs between the various goals and the algorithm's suggested solutions.
- To make the solution more resilient to changes in the issue or the environment, include uncertainty and imprecise information into the optimization process.
- This Algorithm on compilation with Different algorithms will help to optimise it's performance on basis of some weighted parameter which that algorithm uses.

## *2.5 Algorithm Practical Analysis Parameters*

Table 1: Parameters for Analysing Algorithm along with its definition in VM Scheduling

| Parameters | Description |
|---|---|
| Fault Tolerance | It is an ability of an Algorithm to Generate accepting Results even in case of Faults or Errors. VM faults more if the container is not properly virtualised (Build Getting Failed). If the Algorithm has high Fault Tolerance it is not Stable and not Recommended for Scale. |
| Throughput | Total Number of Processes completed by VM in each Time Frame. High Throughput Refers to more Efficiency. |
| Scalability | Ability to handle the increasing number of tasks without much significant decrease in performance. |
| Response Time | Time Referred to respond to the Client from the time request is sent as Process. High Response time refers to less optimised Algorithm. |

## 3.0 RESEARCH GAPS

While there are lot of researches going in area of VM Load Balancing Scheduling Algorithms, there are Gaps which Cloud Providers and Researchers must account upon. Some of the Research Gaps are listed below:

**Heterogeneous environments:** The mostly used VM Scheduling Algorithms are designed with building on-top of same hardware and software dependencies. Heterogeneous environments allow algorithms to dynamically adjust with dividing weightage on parameters itself according to different VM Configurations including their own Performance and Resource Limitations.

**Scalability:** The Technologically invented Algorithms allow for small to Mid-Scale Systems, in short, VMs must tend to horizontal expand its systems so that it can achieve its potential to scale itself on large system.

**Security and Privacy:** VM Scheduling Algorithms often expose their Load Balanced Parameters which can lead to vulnerabilities for security threats. These threats rely on exposing the Client's loaded Requests or Parameters for these Algorithms. Therefore, it is necessary to effectively address the security concerns and ensure the anonymity of the Client/Server Loaded Requests without exposing data and its resources.

**Energy Efficiency:** It is necessary for any algorithm to stand out on Optimise Energy with its other performance metrics. The Cloud Systems rely on Underlying Physical Servers which are powered by electricity. If Algorithms deployed on Systems are not Energy Efficient, it might not be able to achieve scalability, and prioritising Parallel Computing possibilities. The More efficient Algorithms improvise on taking more accurate decisions on Dynamically inputted load which will eventually help to optimise the Algorithm with increase in checkpoints while improving its energy efficiency.

**High computational complexity**, difficulty in handling complex and dynamic workloads, inability to handle non-linear or non-convex optimization problems, slow convergence rates, dependence on monitoring performance metrics that may be time consuming or inaccurate, and the possibility of over-allocation of resources leading to wasted resources and increased costs are all potential issues with VM load balancing scheduling algorithms. Consider their limitations and whether or not they are appropriate for the target cloud computing environment before putting these strategies into practise.

Table 2: Table Showing what parameters from **Table 1** each algorithm should abide to choose them for specific Application [21].

| Algorithm | Fault Tolerance | Throughput | Scalability | Response Time | Migration |
|---|---|---|---|---|---|
| *PSO* [7] | Moderate (Can be Optimised) | Moderate to High | Not Recommended | Low | High |
| *Multi-Objective Optimisation* [4] | Low | High | High | Low | High |
| *Active Monitoring Algorithm* [9,10] | High | High | Low | High | High |
| *Genetic Algorithm* [16] | Moderate | High | Moderate to High | Moderate | Very Low |

## 4.0 CONCLUSION

We've seen Scheduling Algorithms in action. Software architecture must meet algorithmic parameters. Cloud resource scheduling is essential. We employ numerous techniques to reach our cloud aim of appropriate resource utilisation and low computation cost. To get the best results, pick the algorithm that fits your needs. Task scheduling requires scheduler allocation. This scheduling underpins cloud computing load balancing. Hence, the application's job scheduling method reduces execution time. Task-based algorithms help us get the optimum outcome for each task, making machines faster. Some algorithms are less common due to increased computation power or lesser efficiency. Every load balancer maximises resource consumption and distributes tasks in processes to satisfy quick response times for VM Scheduling. VM Scheduling assures distinct instances on Virtualizing platform use maximum resources and load balanced as consumers directly connect with VM instances for feedback and responses. NGINX load balances depending on a threshold load vs. response time. Companies must employ best-fit scheduling algorithms based on paper advantages and disadvantages. Scheduling prepares the architecture and key concept.